\documentclass[sigconf]{acmart}

\usepackage{booktabs} % For formal tables
\usepackage{algorithm}
\usepackage[noend]{algpseudocode}
\usepackage{caption}
\setcopyright{none}
\begin{document}
\title{Parameterizing Kterm Hashing }
%\titlenote{Produces the permission block, and
%  copyright information}
%\subtitle{Extended Abstract}
%\subtitlenote{The full version of the author's guide is available as
%  \texttt{acmart.pdf} document}

\author{Dominik Wurzer}
%\authornote{xxxx}
\orcid{xxx-xxx-xxx}
\affiliation{%
  \institution{School of Information Management \\Wuhan University}
  \streetaddress{P.O. Box 1212}
  \city{ }
  \state{dominik@wurzer.com}
  \postcode{xxx-xxx}
}
\email{ }

\author{Yumeng Qin}
\authornote{Corresponding Author}
\orcid{xxx-xxx-xxx}
\affiliation{%
  \institution{School of Information Management \\Wuhan University}
  \streetaddress{P.O. Box 1212}
  \city{ }
  \state{yumeng.qin@whu.edu.cn}
  \postcode{xxx-xxx}
}
\email{ }


\begin{abstract}
Kterm Hashing provides an innovative approach to novelty detection on massive data streams. Previous research focused on maximizing the efficiency of Kterm Hashing and succeeded in scaling First Story Detection to Twitter-size data stream without sacrificing detection accuracy. In this paper, we focus on improving the effectiveness of Kterm Hashing. Traditionally, all kterms are considered as equally important when calculating a document's degree of novelty with respect to the past. We believe that certain kterms are more important than others and hypothesize that uniform kterm weights are sub-optimal for determining novelty in data streams. To validate our hypothesis, we parameterize Kterm Hashing by assigning weights to kterms based on their characteristics. Our experiments apply Kterm Hashing in a First Story Detection setting and reveal that parameterized Kterm Hashing can surpass state-of-the-art detection accuracy and significantly outperform the uniformly weighted approach.\\
\end{abstract}

\copyrightyear{2018}
\acmYear{2018}
\setcopyright{acmlicensed}
\acmConference[SIGIR '18]{The 41st International ACM SIGIR
Conference on Research and Development in Information
Retrieval}{July 8--12, 2018}{Ann Arbor, MI, USA}
\acmBooktitle{SIGIR '18: The 41st International ACM SIGIR Conference
on Research and Development in Information Retrieval, July 8--12,
2018, Ann Arbor, MI, USA}
\acmPrice{15.00}
\acmDOI{10.1145/3209978.3210101}
\acmISBN{978-1-4503-5657-2/18/07}

\fancyhead{}

\maketitle

\section{Introduction}
Kterm Hashing [18] provides an innovative approach to novelty detection. When applied to streaming tasks, like First Story Detection (FSD), it exceeds the efficiency of state-of-the-art algorithms by several orders of magnitude, without sacrificing effectiveness. Kterm Hashing forms compound terms, called kterms, from all unique terms in a document. The document's degree of novelty is computed by the number of unseen kterms in proportion to the document length. When Kterm Hashing was first introduced, all kterms were considered as equally important for quantifying the degree of novelty. We believe that uniform kterm weights are sub-optimal for tasks like FSD, as kterms like \textit{\{the, is, 23\}} would carry the same weight as the kterm \textit{\{downtown, earthquake, LA\}}. By intuition, the latter one appears to be more helpful to discover new events in data streams. We propose to abandon the principle of uniform kterm importance and instead place weights on kterms to boost detection effectiveness. Learning weights for all kterms is impractical because their number is high\footnote{binominal coefficient of document length and kterm length}. Instead of directly learning weights for each kterm, we learn weights for surrogate clusters. These clusters group kterms based on common characteristics, which allows associating a kterm's importance with the weight corresponding to its nearest cluster. Our experiments in Section 4 show that \textit{parameterized} Kterm Hashing can significantly outperform \textit{uniformly} weighted Kterm Hashing for FSD.

\subsection{Related Work}
First Story Detection (FSD) describes the research task of monitoring a stream of documents with the intent of identifying those documents that speak about previously unknown events first [4]. FSD systems detect new events using a fixed thresholding strategy. This requires the computation of a novelty score for each document. Documents whose novelty score exceeds the detection threshold are considered to speak about new events [3]. The traditional approach to FSD [1,2,3,15] calculates the novelty of a document by its distance to its nearest neighbour. This is known to be the most effective approach in FSD [3] but also among the slowest, as the time complexity depends on the number of comparisons made. \\\\
Kterm Hashing [18] offers a new approach for novelty computation without ``document-level'' comparisons. Instead, Kterm Hashing constructs a single representation of the past - the memory - and compares new documents to it. This provides a single point of comparison which results in a higher efficiency than document-level comparison strategies [14]. 
Kterms are compound terms of length $k$, based on all terms that appear in a document. By kterm length ($k$), we refer to the number of compound terms. Upon arrival of a new document from the stream, Kterm Hashing exhaustively forms all kterms up to length $k$. Novelty is computed by the ratio of unseen kterms with respect to the memory and the number of kterms formed. Newly encountered kterms are subsequently made persistent in the memory for future calculations. The single point of comparison shifts the time complexity from the number of encountered documents to the number of kterms per document, i.e. the binominal coefficient of document and kterm length. The original publication [18] determines the membership of kterms in the memory by hashing them onto a fixed sized Bloom Filter [5] to ensure fast look-ups in constant time and space [14]. They report FSD accuracy on par with UMass [3] and LSH-FSD [15], while processing data streams up to 2 orders of magnitude faster. Our experiments reveal that detection effectiveness can be increased by distinguishing between different levels of kterm importance. \\\\
Our approach of computing a document's novelty based on the novelty of weighted terms is distantly related to the concept applied by the IDF-FSD system [10]. Their novelty estimation relies on the sum of term frequencies and inverse document frequencies with respect to previously encountered documents. Both features are commonly used by various Information Retrieval applications. In addition to FSD, Kterm Hashing was also applied to Rumour Detection [13] to significantly improve detection accuracy.

\section{Method}
Kterm Hashing [18] estimates novelty based on the fraction of unseen kterms in a document. Instead of considering all kterms as equal, we want to distinguish them based on their importance for the novelty computation. \\\\
The number of kterms spawned by a document depends on the document length and kterm length. We choose kterms of length 1 to 3, which were found to perform best for FSD [18]. A document with 10 unique terms\footnote{average length of a tweet} spawns 175 unique kterms for length 1 to 3. Heaps Law [21] states that the vocabulary size grows without bound. Since kterms are formed by exhaustively compounding terms, their number grows faster than the collection vocabulary. The streaming nature of FSD renders individual weights for kterms infeasible. To mitigate this problem, we shift from individual kterm weights to weights for kterm categories.\\
\begin{table}[h]
\centering
\begin{tabular}{|c|c|}
\hline
\textbf{kterm feature} & \textbf{Description}  \\ \hline
inverse document & idf of kterm \\
frequency  &  components \\ \hline
term frequency &   tf of kterm components   \\ \hline
document frequency &  df of kterm components  \\ \hline
entity & 4 different entities  \\ \hline
part of speech  & 4 different POS tags  \\\hline
spelling &   ratio of correctly spelled words  \\ \hline
numbers &  presence and frequency \\
& of numbers   \\ \hline
twitter specific &hashtags and usernames   \\ \hline
kterm length & number of compound terms \\ \hline
\end{tabular}
\caption{Each kterm feature category provides several features based on the number of occurrences, sum, min, max and average of feature values}
\label{tab_eff}
\end{table}
\vspace{-8mm}\subsection{Forming kterm Categories}
When detecting new events, we weigh each kterm based on the weight of its category. These categories enclose kterms that share similar characteristics. Table 1, lists 9 feature categories of which we form 60 kterm features. Algorithm 1 describes the construction of kterm categories and how we associate them with the documents in the training set. Each kterm category is a cluster represented by a centroid vector. In \textit{Line 1}, we initialize the centroid matrix (\textit{CENTROID}) to hold random values for each of the 60 kterm features. We then construct a dense kterm matrix (\textit{KTERM}) by exhaustively forming kterms form all documents in the training set and extract the kterm feature values for each of them (\textit{Line 2}). \textit{Lines 3} to \textit{8} apply K-mean clustering [22] to group the kterms into 120 categories. This requires computing the similarities ($\textnormal{\textit{SIMS}}_{1}$) of the kterm vectors with all centroid vectors using the dot product between their matrices. Kterms are assigned to their nearest cluster by identifying the highest similarity value in each matrix column (\textit{Line 5}). This turns the similarity matrix ($\textnormal{\textit{SIMS}}_{1}$) into a sparse matrix that associates kterms with their corresponding category (cluster). The clusters are then re-computed by updating their centroid vector based on the average of the kterm vectors assigned to it (\textit{Line 6 - 7}). K-mean clustering ensures that kterms with similar feature characteristics are likely to end up in the same cluster. We determined the number of iterations and categories empirically and found that fewer categories reduce the detection accuracy, while more categories only result in marginally better accuracy.
The training set for the subsequent learning procedure is formed by the dot product between the document matrix and $\textnormal{\textit{SIMS}}_{1}$, the matrix associating kterms with their corresponding kterm category (\textit{Line 9}).\\
\begin{algorithm}
\caption{\textbf{: kterm categories}\\\textbf{Input:}\\  DOCS [ documentID $\times$ ktermID ]\\KTERM [ ktermID $\times$ featureID]\\CENTROID [centroidID $\times$ featureID]\\SIMS [centroidID $\times$ ktermID]\\\textbf{Output:}\\TRAININGSET [documentID $\times$ centroidID]}
\label{alg_CMU}
\begin{algorithmic}[1]
\State{CENTROID $\leftarrow$ \textit{random}(kterm features) }
\State{KTERM $\leftarrow$ \textit{kterm features} (DOCS) }
\For{\emph{iteration in $\{1...100\}$}}
	\State{$\textnormal{SIMS}_{1} \leftarrow \textnormal{CENTROID} \bullet \textnormal{KTERM}^{T}$}
	\State{$\textnormal{SIMS}_{1} \leftarrow colmax (\textnormal{SIMS}_{1})$}
	\State{$\textnormal{SIMS}_{2} \leftarrow uniform(\textnormal{SIMS}_{1})$}
	\State{$\textnormal{CENTROID} \leftarrow \textnormal{SIMS}_{2} \bullet \textnormal{KTERM}$}
\EndFor 
\State{\textbf{end}}
\State{$\textnormal{TRAININGSET} \leftarrow \textnormal{DOCS} \bullet \textnormal{SIMS}_{1}^{T}$}
\end{algorithmic}
\end{algorithm}
\subsection{Parameterizing Kterm Hashing }
The previous section grouped kterms into categories and assigned the documents in the training set to them. Before learning optimal kterm category weights, we divide the training set into 2 classes, ``first stories'' and ``follow-ups'', and apply feature scaling to ensure all feature values are within the same range. The kterm category weights are optimized by a Support Vector Machine (SVM) classifier [23], which we found to perform best. The SVM uses a radial basis function kernel ($e^{(-\gamma*|u-v|^2)}$) with gamma ($\gamma$) of $0.15$  and a convergence tolerance of $0.1$. Note that FSD data sets are highly imbalanced because every first story is associated with several follow-ups. Our experiment section (Section 4) compares strategies to counteract this imbalance.\\  
\begin{equation}
\begin{split}
 N(d_n) = \sum_{kt \in d_n} \alpha_{kt} \dbinom{|d_n|}{|kt|}^{-1} 
\begin{Bmatrix} 1 : kt {\not\in} M_{n-1} \\ 0 : kt {\in} M_{n-1} \end{Bmatrix}
\end{split}
\label{eq:novKtermFinal}
\end{equation}
\\At run-time, we incorporate our kterm category weight ($\alpha_{kt}$) into the equation of Kterm Hashing (Equation 1). The novelty of document $d_n$, is based on the kterm category weights ($\alpha_{kt}$) of its kterms ($kt \in d_n$) if they are new with respect to the memory of the past $M_{n-1}$.
\section{Experiments}
In a streaming setting, like FSD, documents arrive on a continual basis one at a time [19]. We require our approach to compute a novelty score for each document in a single-pass over the data. To evaluate the accuracy of our parameterized version of Kterm Hashing, we compare it to the traditional approach on two massive Twitter FSD data sets. %Messages with high rumour score are considered likely being rumours. The classification decision is based on an optimal thresholding strategy using cross-validation. We report accuracy to evaluate effectiveness, as is usual in the literature (Zhou et. al, 2015). Additionally, we evaluate the efficiency of our approach, measured by the throughput per second, when applied to a high number of messages. %Additionally we use the standard TDT evaluation procedure (Allan, 2002) with the official TDT3 evaluation scripts by NIST(xxx link in footnote) using standard settings. This procedure evaluates detection tasks using Detection Error Trade-off (DET) curves, which show the trade-off between miss and false alarm probability. By visualizing the full range of thresholds, DET plots provide a more comprehensive illustration of the effectiveness than single values metrics (Wurzer et. al, 2015).

%xxx show table with translated positive and negative examples
\begin{table}[b]
\centering
\begin{tabular}{|c|c|c|}
\hline
\textbf{Kterm Hashing} & $\boldsymbol{C_{min}}$ & \textbf{Difference (\%)}   \\ \hline
traditional & 0.8021 & -\\ \hline
pkterm hashing & 0.7996  & -0.31\%  \\ \hline
pkterm hashing + & & \\
class weight & 0.7896 & -1.31\% \\ \hline
pkterm hashing + & & \\
 class weight + & \textbf{0.7822}  & -2.48\% \\ 
  skip evaluation & & \\ \hline
\end{tabular}
\caption{Effectiveness of different variants of Kterm Hashing on the ``\textit{Cross-Twitter}'' data set.}
\label{tab_eff}
\end{table}

\begin{table}[b]
\centering
\begin{tabular}{|c|c|c|}
\hline
\textbf{Kterm Hashing} & $\boldsymbol{C_{min}}$ & \textbf{Difference (\%)}   \\ \hline
traditional & 0.7721 & -\\ \hline
pkterm hashing & 0.7689  & -0.31\%  \\ \hline
pkterm hashing + & & \\
class weight & 0.7456 & -3.43\% \\ \hline
pkterm hashing + & & \\
 class weight + & \textbf{0.7378*}  & -4.44\% \\ 
  skip evaluation & & \\ \hline
\end{tabular}
\caption{Effectiveness of different variants of Kterm Hashing on the ``\textit{Large-scale Twitter Corpus}'' data set. Asterisk (*) indicates statistically significant differences $(p<0.05)$.}
\label{tab_eff}
\end{table}

\begin{table*}[]
\centering
\begin{center}
\begin{tabular}{|c|c|c|c|c|c|c|c|c|} \hline
feature:  & \textbf{idf} & \textbf{tf} &  \textbf{df} & \textbf{entity} & \textbf{POS}& \textbf{spelling}& \textbf{Twitter}& \textbf{length}  \\ \hline
relative impact    & & & & &&&& \\ 
on detection & -4.15\%  & -2.69\% & -3.72\% &  -\textbf{5.23}\% &  -3.54\% &  -4.22\% &  -2.55\% &  -4.92\% \\ 
cost ($C_{min}$):  & & & & &&&& \\ \hline
\end{tabular}
\captionsetup{justification=centering}
\caption{Features ablation: impact on performance when removing a feature group.\\  \textit{idf}: inverse document frequency; \textit{tf}: term frequency; \textit{df}: document frequency; \textit{POS}: part of speech;}
\end{center}
\end{table*}

\subsection{Data Set}
The first data set, ``Cross-Twitter''\footnote{The Cross Project is a joint venture between the University of Edinburgh and the University of Glasgow, http://demeter.inf.ed.ac.uk/cross/}, was also used in the original Kterm Hashing paper [18]. It consists of 27 topics and 115,000 tweets ordered by their publication time-stamp. Additionally, we use the 500 topics and 150,000 tweets of the ``Large-scale Twitter Corpus'' [11]. Both data sets are frequently used to evaluate the performance of FSD systems [9, 11, 12, 15, 18, 19]. %The 500 topics and corresponding annotated tweets are randomly split in half to provide a training and test set. This leaves us with a 250 topics training set and two tests sets spanning 250 and 27 topics.
\subsection{Evaluation Metric}
As is common in Topic Detection and Tracking [4], we evaluate FSD accuracy by the normalized Topic Weighted Minimum Detection Cost, dubbed $C_{min}$. The detection cost $C_{min}$ provides a linear combination of miss and false alarm probabilities, which allows comparing different methods based on a single value metric [2]. We make use of the standard TDT evaluation procedure [4] with the official TDT3 evaluation scripts using the standard settings.

\subsection{Improving the accuracy of Kterm Hashing for FSD}
Table 2 compares the effectiveness of parameterized Kterm Hashing - dubbed pkterm hashing - with traditional Kterm Hashing on the ``Cross-Twitter'' data set, which was also used by the original paper [18]. Following the original publication, we determine the kterm length parameter for traditional Kterm Hashing using grid search. For this experiment all systems make use of the 500 topics from the ``Large-scale Twitter Corpus'' as training data for parameter optimization. In contrast to our expectations, Table 2 shows only a marginally better detection cost for pkterm hashing. Deeper analysis of the training procedure revealed that the potential of pkterm hashing is limited by the class imbalance of the training data.\\\\
\textbf{Counteracting class imbalance by class weights:}\\
In FSD, each first story (detection target) is usually followed by several follow-ups, creating a class imbalance for the training data of the learning algorithm. Classical sampling methods [8] harm the detection accuracy because the imbalance exceeds 1:1,000. We address this imbalance by placing a class weight (Equation 2) on the training set [7]. This increases the importance of the ``detection target'' class when learning kterm category weights using a Support Vector Machine. Table 2 shows that class weights successfully improve the detection accuracy of parameterized Kterm Hashing by 1.3\% in comparison with traditional uniformly weighted kterms.\\
\begin{equation}
\delta_{class A}(\frac{\#\textnormal{instances in class A}}{\#\textnormal{instances in all classes}}*0.3)
\end{equation}\\
\textbf{Increasing effectiveness by Skip Evaluation:}\\
Machine learning algorithms tend to produce better results when exposed to more training data [17]. Skip Evaluation is a frequently applied methods in the Topic Detection and Tracking (TDT) program [4] to increase the number of topics in a data set without annotating new topics [1, 20]. Skip Evaluation iterates a certain number of passes over each topic. At each pass, the first story (detection target) of each topic is removed (skipped), making the first follow up document the new detection target. This doubles the number of detection targets and reduces the number of follow-ups by one per pass. We limited Skip Evaluation to 10 rounds on the training set to prevent small scale topics from vanishing. Skip Evaluation by itself does not resolve the class-imbalance of the training set, which requires additional class weights. Note that Skip Evaluation is only carried out on the training set to determine optimal parameter weights and \textit{not} on the test set for Kterm Hashing. Table 2, shows that parameterized Kterm Hashing benefits from the increased training data size, as it outperforms the traditional approach by 2.48\%. Although the improvement might appear minor, we want to point out that detection accuracy surpasses the reported effectiveness [18] of the UMass FSD system [3], which is considered to be the state-of-the-art in terms of detection accuracy [14].\\\\
We repeated the experiments on the ``Large-scale Twitter Corpus'', as seen in Table 3. Unfortunately, ``Cross-Twitter'' provides insufficient training examples (27 topics) to serve as a training set. Therefore, we randomly split the 500 topics of the ``Large-scale Twitter Corpus'' to create a training and test set with 250 topics each. Table 3 confirms the findings of  increased detection accuracy of parameterized Kterm Hashing in conjunction with class weights and skip evaluation. Following the higher number of topics, the difference in detection cost reaches statistical significance $(p<0.05)$.
\vspace{-1.5mm}\subsection{Feature Analysis}
The previous experiments confirmed that uniform kterm weights are sub-optimal for FSD. Next, we analyze the impact of certain kterm features on the detection accuracy of parameterized Kterm Hashing. Kterms are weighed based on the weight of their category. When analyzing features, we focused on the kterm features that determine the kterm category, instead of analyzing the category weight itself. To analyze a feature's impact on the detection cost, we apply feature ablation. Feature ablation measures the relative change in detection cost when applying all but one feature [13]. In our case, feature ablation measures the impact on $C_{min}$ by excluding feature when forming kterm categories, as seen in Table 4. The table reveals that features based on entities, inverse document frequency and Part of Speech are particularly useful kterm features. To our surprise, Twitter specific features, like hashtags, appear to have a minor impact on detection cost. We further investigated the impact of hashtags on Kterm Hashing by removing them from the training and test set and measure a 3.68\% relative reduction in detection cost. This is interesting as several approaches [6,16] for First Story Detection rely on hashtags. Manually inspection of annotated topics revealed that the majority $(>60\%)$ of hashtags does not occur in the first story, but in the follow-ups and 6 out of 27 Cross-Twitter topics don't contain any hashtags. Since hashtags are often previously unseen terms, they spawn a high number of unseen kterms, which increase the novelty of followups and decreases detection accuracy.
\vspace{-1.5mm}\section{Conclusion}
Traditional Kterm Hashing considers all kterms as equally important when calculating novelty on data streams. We showed that uniform kterm weights are sub-optimal for FSD on two separate data sets. Instead of placing individual weights on kterms, we group them into categories and learn optimal weight settings for them. Our experiments demonstrated how parameterized Kterm Hashing can significantly outperform the traditional approach for FSD. We also demonstrated that parameterized Kterm Hashing in conjunction with class weights and sufficient training data, can outperform state-of-the-art FSD systems in terms of detection accuracy.

\bibliographystyle{ACM-Reference-Format}
%\bibliography{sample-bibliography}

\end{document}